# ELEMENTS OF VALIDATION OF ARTIFICIAL LIGHTING THROUGH THE SOFTWARE CODYRUN: APPLICATION TO A TEST CASE OF THE INTERNATIONAL COMMISSION ON ILLUMINATION (CIE)

Fakra A. H., Miranville F., Dimitri B., Boyer H.

Physics and Mathematical Engineering for Energy and Environment Laboratory (*e*PIMENT)
University of Reunion
17, Rue du Général Ailleret, 97430 Le Tampon, La Réunion, France
Corresponding email: fakra@univ-reunion.fr

**ABSTRACT**
CODYRUN is a software for computational aeraulic and thermal simulation in buildings developed by the Laboratory of Building Physics and Systems (L.P.B.S). Numerical simulation codes of artificial lighting have been introduced to extend the tool capacity. These calculation codes are able to predict the amount of light received by any point of a given working plane and from one or more sources installed on the ceiling of the room. The model used for these calculations is original and semi-detailed (simplified). The test case references of the task-3 TC-33 International Commission on Illumination (CIE) were applied to the software to ensure reliability to properly handle this photometric aspect. This allowed having a precise idea about the reliability of the results of numerical simulations.

## 1. Introduction

It is important to check the degree of software reliability to simulate any particular physical phenomenon. This allows the detection of limits and potentials of the simulation code and put in evidence indications to improve it. For all these reasons we have initiated a validation study of artificial lighting with the software CODYRUN [1]. The scientific literature has brought to light more closely to the task TC-3-33 ICE [2]. Indeed, in its report, the Committee has developed procedures and test cases to study the reliability of a software simulation of lighting. In this paper, the reference test cases developed (scenarios 1 and 3) are presented, then simulation results obtained when these tests are applied to CODYRUN software are shown.

## 2. New simplified model for calculating indoor lighting

Many studies have been conducted to quantify indoor artificial lighting [3]. The category of numerical models such as Radiosity or Ray tracing can be applied in the context of qualitative study of artificial lighting [4]. CODYRUN to determine indoor lighting quantitatively, from several combined models that take into account the part of diffuse and direct artificial light. The new simplified model, which was developed during this study, is similar to those that have been introduced into software such as DIALUX [5] or CALCULUX [6].

**2.1 Hypothesis of the simulation in CODYRUN**
The light scattering is considered Lambertian (light propagates uniformly along all possible directions). The manufacturer gives the photometric light sheet. The position of the light source in space is know from the Cartesian coordinates of its centre of gravity S. all luminary (sources) or other large area, namely point (plane), cylindrical, and spherical, will be reduced to a point (centre of gravity of the source). No obstruction is located between the illuminated point and the light source.

**2.2 Direct part of lighting (from artificial light source)**
Direct lighting calculation from artificial light depends strongly on the morphological structure (the form) of the light source and the solid angle relative to the illuminated point. Figure 1 shows the direction of propagation of direct part of light on a point of work plane, from a source mounted on the ceiling. The relationship gives the value of direct light, in this case, given by (1).

**2.3. Diffuse part of lighting (from indoor inter-reflection)**
The direct part of lighting is reflected on the interiors surfaces and thus produces a diffuse part of lighting from inter – reflection. These inter – reflection depends on the colour of the walls and therefore the degree of reflectivity of the latter (Figure 2). The model used in CODYRUN takes into account these colours. Indeed, the part of diffuse lighting is calculated by weighting the value of direct part of lighting of the average reflection coefficient of internal walls (specific to each colour). Expression of the calculating diffuse lighting from these inter – reflection is given by the relation (1).

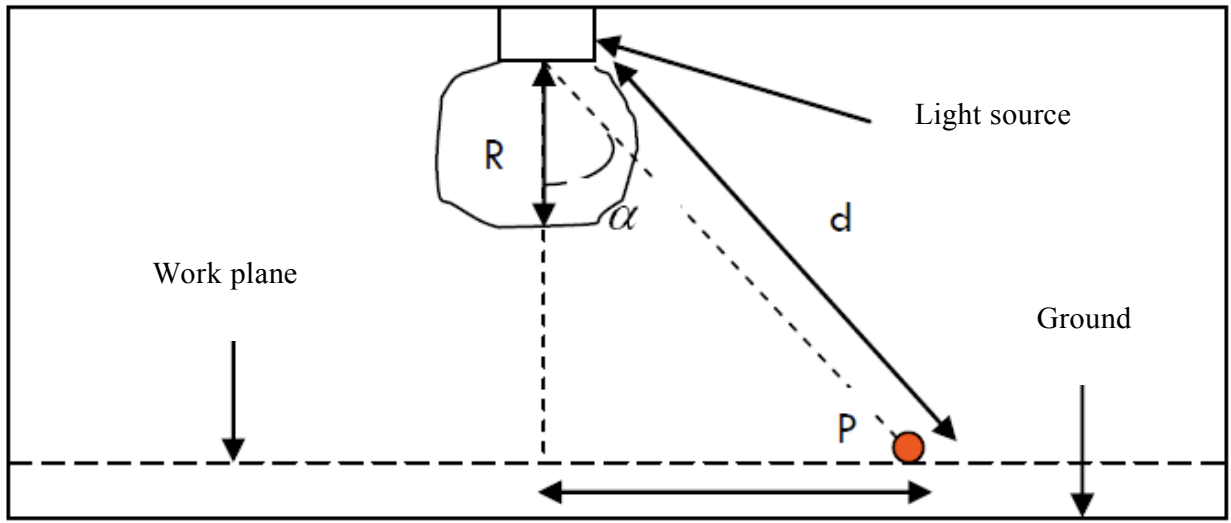

**Figure 1: Direct light source**

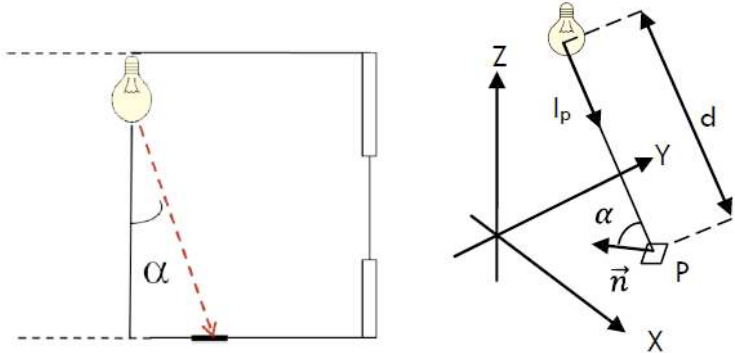

**Figure 2: Diffuse light source (inter – reflection)**

### 2.4. Global part of lighting (combination of simplified models - new approach to modelling)

Those who were integrated into our type of software are simplified. Indeed, we have implemented a very simple relation to determine the artificial illumination from one or more lights reaching any point of a useful plan. The formula is given below:

$$E_{glob.p} = \left(\frac{I_p}{d^2} \times \cos\alpha \times \rho_{moy.}\right) + \left(\frac{I_p}{d^2} \times \cos\alpha\right) = \left(\frac{I_p}{d^2} \times \cos\alpha\right)(1 + \rho_{moy})$$

Where :
- $I_p$ : Luminous intensity of the source in the direction of the illuminated point p (Cd)

- $\alpha$ : Angle between the normal to the plane containing the illuminated point and the line jointing the source to the illuminated point p (°)
- d : Distance between the illumined point p and source light S (m).
- $\rho_{moy}$ : average reflection coefficient (%)

### 3. Presentation of C.I.E tests cases: scenario 1 and 3

#### 3.1. Local geometry
The local test is rectangular and the dimensions are 6.78 m by 6.72 m. The ceiling height is 3.24 m.

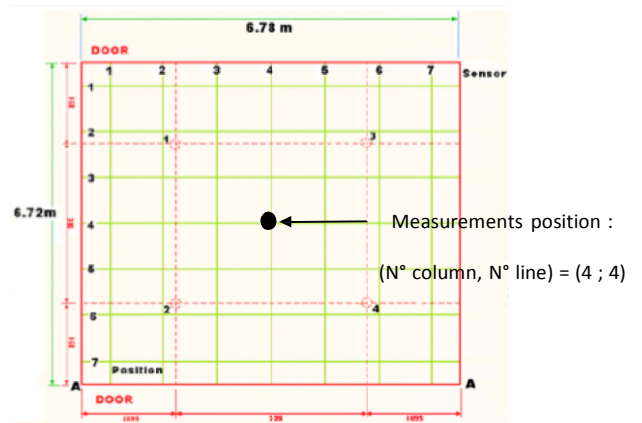

**Figure 3: Ground dimensions of building test CIBSE (artificial lighting)**

#### 3.2. Photometric of light sources
The test building consists of 4 types of fluorescent Philips lamps distributed as shown in the Figure 3 above as four points 1, 2, 3 and 4 of the ceiling. Two configurations (configuration 1 and configuration 3) were studied.

Depending on the type of configuration, the lights were respectively the coordinates and characteristics of the following tables :

- Configuration 1 (CFL, gray wall)

| Light source | Coordinates (m) | Luminous flux (lm) | Luminous intensity (Cd) |
|---|---|---|---|
| 1 | (1.695 ; 1.680; 3.14) | 2182 | 694,55 |
| 2 | (1.695 ; 5.040 ; 3.14) | 2196 | 699,1 |
| 3 | (5.085 ; 1.680 ; 3.14) | 2203 | 701,236 |
| 4 | (5.085 ; 5.040 ; 3.14) | 2182 | 694,55 |

**Table 1: Light source photometric for configuration 1 to the task TC3-33 (CIE)**

The reflection coefficients of the internal walls are all equal to 41% + / - 0.02% for configuration 1 and the lighting is rated to 3.14 m.

- Configuration 2 (semi-specular reflection; gray wall)

| Light source | Luminous flux (lm) | Luminous intensity (Cd) |
|---|---|---|
| 1 | 4087.7 | 1301 |
| 2 | 4174.7 | 1328 |
| 3 | 4135.0 | 1316 |
| 4 | 4114.3 | 1309 |

**Table 2: Light source photometric for configuration 3 to the task TC3-33 (CIE)**

The reflection coefficients of internal walls are 52% + / - 0.02% for scenario 3. The lights sources have same rating as before.

The properties of reflection coefficients are described in the table below :

| Walls | Reflection coefficients | Absorption coefficients |
|---|---|---|
| Ceiling | 0.70 | 0.30 |
| Floor | 0.06 | 0.94 |
| Vertical Wall | 0.41ou 0.52 | 0.59 or 0.48 |

**Table 3: Reflectance of internal walls of the task TC3-33**

The working plane of the grid surface measurement is 0.8 m height above ground. The 7x7m grid contains surface points distant from each other by 0.9 m. The gap between the gate and wall surface constituting the room is 0.48 m.

### 3.3. References measurements

Two error values ($E_m$ and $E_g$) were provided by CIBSE [7]. One concerns the errors due to the measurements noted $E_m$, and another noted $E_g$ is due to all the errors (measurement and simulation) tolerable.

Table 4 show the 4 possible values of errors.

The simulation values are compared to those of lower and upper limits of measurements (including nearby measurements + / - average error then measurements + / - global error) to illuminance at a point. The average illumination in the room will, in turn, be bounded by the measurements + / - Global errors.

## 4. Applications test cases and results of configuration 1

Only the base case configuration 1 will be presented. It is very important to emphasize that measurements do not take into account the phenomenon of bi-directional light source. This means not taking into account the indirect part of the transmitted light within the diffuse internal component in the simulation.

Only a comparison of results obtained in one case (first line of our measurement points) among 6 other tables of values of simulations will be shown. In this case, the reference points considered are represented by (1;1) to (1;7) (refer to Figure 4 for an example of identification of a measuring point of the test case). The curves LS-Eg and LI-Eg are respectively the top and bottom margins of the global reference values (measured values + measurements errors + simulation errors). While the curves LS-Em and LI-Em represent respectively the upper and lower limits of references values averages (measured values + measurements error).

The overall results of comparison between simulation software and reference scenario 1, obtained are given in Appendix. The maximum error was committed in (6;6) (refer to Appendix) and is equal to approximately 15.44 % compared to the overcall upper limit of reference (LS-Eg). The minimum error is approximately 0.71 % (at position (3;3)) compared to the upper limit of reference (LS-Eg). The average relative error is 6.1 %.

|  | Type | |
| --- | --- | --- |
| Error | Point | Average |
| average error ($E_m$) | +/− 3.8% | +/− 6.3% |
| Global error ($E_g$) | +/− 6.7% | +/− 10.5% |

Table 4: Errors of the configurations 1 and 3 of task TC3-33 (CIE)

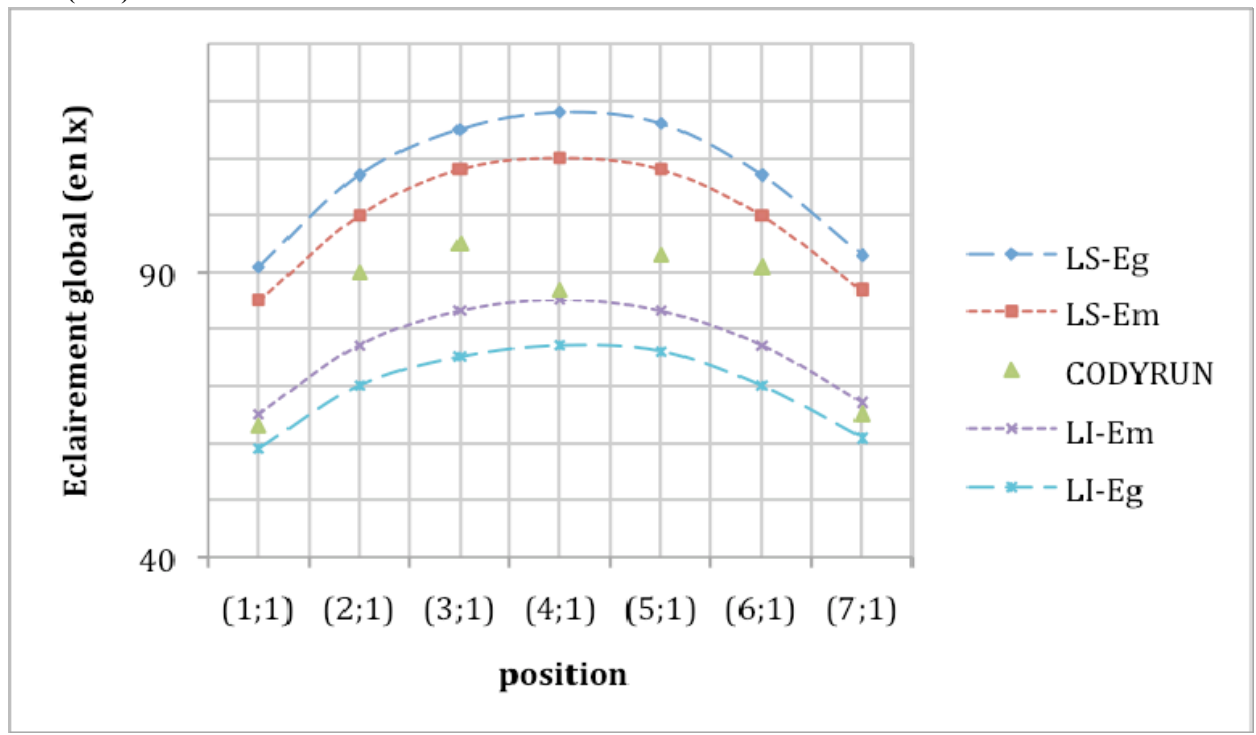

Figure 4.: Configuration 1 of task TC3-33 (CIE) : illuminance at measured points.

The average illuminance is given by the table below:

|  | average illuminance (in lx) |
| --- | --- |
| **Upper limit** | 112 |
| **CODYRUN** | **112** |
| **Lower limit** | 88 |

Table 5: Reference value for the scenario 1 of task TC3-33 (CIE) : average illuminance

## 5. Conclusions

Simulations tests under CODYRUN showed that 29 points on 49 have values between the upper (LS-Em) and bottom (LI-Em) reference, then only 34 points out of 49 have values between the two margins overall limits (above LS-Eg and lower LI-Eg). In both cases, there are more points inside the boundary curves as points beyond. This gives an indication of the reliability of software to simulate artificial interior lighting. In fact, CODYRUN has a reliability of approximately 69 % in this case (for indication, DIALUX 4.0 has obtained a reliability of 75 % for the same test case [8]).